\documentclass[reprint,
 amsmath,amssymb,
 aps,
showkeys,
showpacs
]{revtex4-2}

\usepackage{slashed}
\usepackage{braket}
\usepackage{xcolor}

\usepackage{comment}
\usepackage{graphicx}
\usepackage{dcolumn}
\usepackage{bm}

\begin{document}

\preprint{MUPB/Conference section: }

\title{Generalized LKF transformations for $N$-point fermion correlators in QED \\}

\author{\underline{José Nicasio}}
 \email{jose.nicasio@umich.mx (Corresponding author)}
\affiliation{%
Instituto de Física y Matemáticas,\\ Universidad Michoacana de San Nicolás de Hidalgo, Morelia, México
}

\author{Naser Ahmadiniaz}
 \email{n.ahmadiniaz@hzdr.de}

\affiliation{
 Helmholtz-Zentrum Dresden-Rossendorf,\\
 Bautzner Landstra\ss e 400, 01328 Dresden, Gernamy
}%

\author{James P. Edwards}
 \email{jpedwards@cantab.net}
 \affiliation{Present address, Centre for Mathematical Sciences, University of Plymouth, Plymouth, PL4 8AA, UK.}

\author{Christian Schubert}
 \email{christianschubert137@gmail.com}
 \affiliation{%
Instituto de Física y Matemáticas,\\ Universidad Michoacana de San Nicolás de Hidalgo, Morelia, México
}

\date{\today}

\begin{abstract}
\noindent Within the worldline approach to quantum electrodynamics (QED), a change of the photon's covariant gauge parameter $\xi$ is investigated to analyse the non-perturbative gauge dependence of the configuration space fermion correlation functions, deriving a generalization of the Landau-Kalatnikov-Fradkin transformations (LKFt). These transformations reveal how the non-perturbative gauge dependence of position space amplitudes can be absorbed into a multiplicative exponential factor. 
\end{abstract}

\maketitle


\section{Introduction}\label{intro}
\noindent The LKFt, derived by Landau and Khalatnikov \cite{LandauLKFT}, and independently by Fradkin \cite{FradkinLKFT}, are non-perturbative transformations which relate field theory propagators (or vertices) calculated in different covariant gauges, originally phrased in position space. The transformation was derived first in QED and then generalized to correlation functions in Quantum Chromodynamics (QCD) \cite{LKFQCD,DALLOLIO2021115606}. In particular cases, as in scalar QED in 3-dimensions, it is possible to obtain exact LKFt in momentum space  \cite{Villanueva-Sandoval:2013opv}.

Their applications range from non-perturbative analyses, implying constraints for a non-perturbative ansatz for the fermion-boson vertex \cite{Burden} in the context of the Schwinger-Dyson equations; to perturbative ones, such as to obtain information on Feynman diagrams at higher loop orders, starting from an amplitude at a fixed loop level \cite{DALLOLIO2021115606, AdnanRaya}. The extra information at higher loops is always gauge dependent, and is such that the propagator at fixed loop order calculated in some gauge (e.g. Landau gauge, with gauge parameter $\xi =0$), determined to order $\mathcal{O}(\alpha ^n)$ in the fine structure constant, fixes -- through the LKFt -- the coefficients of the contributions at order $\alpha ^{i+j} \xi ^i$, with $i=0,1,..., \ j=0,1,...,n$. \cite{BASHIR2005259}.

In 2016, a generalized LKFt was derived for the case of $N(=2n)$-point correlator functions \cite{LKFTWL1} in scalar QED, applying the \textit{worldline formalism} (for reviews see \cite{ChrisRev, UsRev}), an alternative formulation of quantum field theory based on first quantised particle path integrals \cite{Strass1}. In this contribution we recap more recent work extending and improving this to spinor QED first presented in \cite{PhysRevD.104.025014}.

\section{The dressed propagator}

\noindent In section II of \cite{NotasJames}, also in these proceedings, the worldline representation of the fermion propagator, ${S^{x'x} = \big\langle x' \big|\big[m - i\slashed{D} \big]^{-1} \big| x\big\rangle}$, in an electromagnetic field, $A_\mu$ is discussed -- see also \cite{fppaper1, fppaper2}. In the covariant derivative $D_{\mu} = \partial_{\mu} + ieA_{\mu}$ the field is specialized to plane waves, 
\begin{align}
A^{\gamma}_{\mu}(x)= \sum ^N_i \varepsilon _{i \mu} \text{e}^{ik_i \cdot x},
\end{align}
to describe external photons scattering off the fermion line, with definite momenta, $k_i$, and polarizations, $\varepsilon _{i \mu}$.

 \subsection{supersymmetric invariance in the wordline}
 \noindent Noting that the worldline action $S[x,\psi,A^{\gamma}]$ described in (3) of \cite{NotasJames} is invariant under the supersymmetric transformations on the worldline (for Grassmann variable $\zeta$)
 \begin{align}
 \delta x^\mu = - 2 \zeta \psi ^\mu, \ \ \ \ \delta \psi ^\mu = \zeta \dot{x}^\mu,
 \end{align}
we can rewrite that action in a more compact form in terms of the superfield and superderivative
 \begin{align}
\hspace{-0.75em} \mathbb{X}^\mu(\tau,\theta)=x^\mu (\tau)& + \sqrt{2} \theta (\psi ^\mu (\tau) + \eta ^\mu)\,, \quad \mathbb{D} = \partial _\theta - \theta \partial _\tau.
 \end{align}
 Here $\theta$ is a Grassmann parameter which extends the parameter domain, $\tau \rightarrow \tau| \theta$. Then the action is written:
 \begin{align}
  S&[x, \psi,A^\gamma]= \int _0 ^T d\tau  \int d \theta \Big[-\frac{1}{4} \mathbb{X} \cdot \mathbb{D}^3 \mathbb{X} - i e A[\mathbb{X}] \cdot \mathbb{D} \mathbb{X} \Big] . \label{S[X]}
 \end{align}

\subsection{Interaction with virtual photons}

\noindent The interaction with a quantum field, $\bar{A}_{\mu}$, can be included by splitting the field $A = A^{\gamma} + \bar{A}$, and integrating over the field $\bar{A}$ (the background field method \cite{Abbott1})
\begin{align}
Z_{\bar{A}}= \int \mathcal{D} \bar{A}\, \text{e}^{-\int d^D x (- \frac{1}{4}\bar{F}_{\mu \nu} \bar{F}^{\mu \nu})-S_{\text{gf}}(\xi)},
\end{align}
with $A^{\gamma}$ the external photon field and where  $S_{\text{gf}}(\xi)$ is the gauge-fixing action depending on gauge parameter $\xi$
\begin{align}
     S_{\text{gf}}=- \int d^4x \, \frac{(\partial \cdot \bar{A})^2}{2 \xi}.
\end{align}
 
The inclusion of the quantum field makes the propagator in the covariant gauge $\xi$  become
 \begin{align}
 \hspace{-1em}&\bar{S}^{x'x}(\xi)=   \Big[m + i \slashed{D}' + i \frac{\delta}{\delta \slashed{J}[x']}\Big] \Big\langle \text{e}^{ie \int d^Dx J[x] \cdot \bar{A}[x]}K^{x'x}\Big\rangle_{\bar{A},\xi} \Big| _{J=0} \nonumber \\
 \hspace{-1em}&= \Big[ m + i \slashed{D}' + i \frac{\delta}{\delta \slashed{J}[x']} \Big] 2^{- \frac{D}{2}} \text{symb}^{-1} \int _0 ^\infty d T \text{e}^{-m^2T} \nonumber \\
 \hspace{-1em}&  \times \int _{x(0)=x}^{x(T)=x'} \mathcal{D} x(\tau) \int _{\text{APC}} \mathcal{D} \psi (\tau) \text{e}^{-S[x,\psi,A^\gamma ] - S_i} \Big| _{J=0},
 \end{align}
 where the insertion of $\bar{A}$ in the pre-factor of the kernel, in (1) of \cite{NotasJames}, is now generated by the functional derivative of the source,  $J[x]$, in $\text{e}^{ie \int d^Dx J[x] \cdot \bar{A}[x]}$.  After evaluating the expression on $J=0$, the term $S_i$ represents the interaction with virtual photons, and is given by
 \begin{align}
 &S_i= \frac{e^2}{2} \iint d^Dy d^Dy' \mathcal{J}(y) \cdot G(y-y';\xi) \cdot \mathcal{J} (y'), \label{Si} \\ 
 & \ \ \ \ \ \mathcal{J}= J^\mu (y) + \int _0^T d \tau \int d \theta \, \delta ^D(y-\mathbb{X}) \mathbb{D} \mathbb{X},
 \end{align} 
where $G_{\mu \nu}(y ; \xi)$ is the configuration space photon propagator in $D$ dimensions in covariant gauge $\xi$. 

\vspace{-0.75em}
\section{N-point fermion correlator}
\noindent The ${N(=2n)}$-point correlation function with covariant gauge parameter $\xi$, defined by  ${\mathcal{S}(x_1 \cdots x_n; x'_1 \cdots x'_n | \xi)=\braket{\bar{\psi}(x_1) \cdots \bar{\psi}(x_n) \psi (x'_1) \cdots \psi (x'_n)}}$ relates to the partial one, $\mathcal{S}_{\pi}$, by summing over all pairings (contractions) of the fields $\bar{\psi}(x_i)$ with the $\psi(x'_j)$:  
 \begin{align}
 \hspace{-0.75em}\mathcal{S}(x_1 \dots x_n;x'_1 \dots x'_n|\xi)= \sum _{\pi \in S_n} \mathcal{S}_{\pi}(x_1 \dots x_n; x'_{\pi (1)} \dots x'_{\pi(n)} | \xi),
 \end{align}
where the fermion lines in $\mathcal{S}_\pi$ go from $x_i$ to $x'_{\pi(i)}$.
The partial $N$-point function in the presence of an electromagnetic field, $A_\mu ^{\gamma}$, and including virtual photons, is  
\begin{align}
\mathcal{S}_\pi (x_1 \dots x_n; x_{\pi (1)} \dots x_{\pi (n)} | \xi)= \Big\langle\prod ^n_{i=1} S_i^{x'_{\pi (i)} x_i} \Big\rangle_{\bar{A}, \xi}, \label{2Nfunction}
\end{align}
with the fermion propagator, $\mathcal{S}_i^{x'_{\pi (i)} x_i}$, as in (1) of \cite{NotasJames}, but now with $A^{\gamma}_\mu + \bar{A}_{\mu}$ instead of just $A^{\gamma}_{\mu}$. The worldline representation of the $N$-point partial amplitude is 
\begin{align}
\hspace{-0.5em}&\mathcal{S}_{\pi}(x_1 \dots x_n; x'_{\pi (1)} \dots x'_{\pi(n)} | \xi) = \prod _{j=1}^n \Big( [m + i \slashed{D}'_j + i \frac{\delta}{\delta \slashed{J}_j[x'_j]}] \Big) \nonumber \\
\hspace{-0.5em}&  \times  \braket{\text{e}^{ie \int d^Dx J[x] \cdot \bar{A}[x]} \prod _{j=1}^n \Big( K^{x'_{\pi(j)}x_j} \Big)}_{\bar{A},\xi} \Big| _{J=0}  \\
\hspace{-0.5em}&= \prod _{j=1}^n \Big( [m + i \slashed{D}'_j + i \frac{\delta}{\delta \slashed{J}_j[x'_j]}] \Big) \prod _{j=1}^n 2^{- \frac{D}{2}} \text{symb}^{-1} \int _0 ^\infty d {T_j} \text{e}^{-m^2T_j} \nonumber \\
\hspace{-0.5em} & \hspace{0.5em}  \int _{x_j(0)=x_j}^{x_j(T_j)=x_j'} \mathcal{D} x_j(\tau_j) \int _{\text{APC}} \mathcal{D} \psi _j (\tau _j) \text{e}^{- \sum _{l=1}^n S^{(l)}[\mathbb{X}_l] - S_{i,n}} \Big| _{J=0},
\end{align}
where $S^{(l)}[\mathbb{X}_l]$ is the action (\ref{S[X]}) of the fermion line $l$ and $S_{i,n}$ is as expressed in (\ref{Si}), but with a sum over lines
\begin{align}
\hspace{-0.75em}\mathcal{J}^\mu (y)= J^\mu (y) + \sum _{l=1}^n \int _0 ^{T_l} d \tau _l \int d \theta _ l \delta ^D (y- \mathbb{X}_l) \mathbb{D}_l \mathbb{X}_l^\mu. 
\end{align}
where we used (2) of \cite{NotasJames} for the kernel $K^{x'_{\pi(i)}x_i}[k_1, \varepsilon_1; \cdots ; k_N, \varepsilon _N | A^{\gamma} + \bar{A}]$.
 
\vspace{-0.75em}
\section{\label{sec:level1}The Generalized LKFt in spinor QED \protect }

\noindent  Interactions with external photons can be expressed in terms of the vertex operator in (5) of \cite{NotasJames}. Then, a gauge transformation can be done by making the replacement $\varepsilon_{\mu} \rightarrow \varepsilon_{\mu} + \xi k_{\mu}$ in the vertex operator and the on-shell invariance of the amplitude is well understood by the Ward-Takahashi identity. So the non-trivial transformation properties of the $N$-point correlation functions only require analysis of the gauge transformation of virtual photons under a variation in gauge parameter $\xi$.

A gauge transformation of virtual photons can be realised by sending $\xi \rightarrow \xi + \Delta \xi$ in the photon propagator $G_{\mu\nu}(y ; \xi)$. The transformation properties of the $N$-point correlation functions are then determined by how $S_{i,n}$ transforms under this change. The derivation of this generalized LKFt for spinor QED can be found in \cite{PhysRevD.104.025014}.   Essentially, the idea is to analyse the expectation value of the product of kernels, $\braket{\prod _{i=1}^n K^{x'_{\pi (i)} x_i}}_{\bar{A},\xi}$ given by
 \begin{align}
 &\prod _{j=1}^n 2^{- \frac{D}{2}} \text{symb}^{-1} \int _0 ^\infty dT_j \text{e}^{-m^2 T_j} \int _{x_j(0)=x_j}^{x_j(T)=x'_j} \mathcal{D}x_j(\tau _j)  \nonumber \\
 & \times \int _{\text{APC}} \mathcal{D} \psi _j(\tau _j) \text{e}^{- \sum _{l=1}^n S^{(l)}[x_l,\psi _l,A _l^{\gamma}]-\sum _{k,l=1}^nS_{i\pi}^{(k,l)}}, 
 \end{align}
with $S_{i \pi}^{(k,l)}$ as defined in (\ref{Si}), but with $J=0$. Its variation due to a gauge transformation of virtual photons is
\begin{align}
&\Delta _{\xi} S_{i \pi} ^{(k,l)} = \Delta \xi \frac{e^2}{32 \pi ^{\frac{D}{2}}} \Gamma \left(  \frac{D}{2} - 2   \right) \left\{[(x_k - x_l)^2]^{2- D/2} \right. \nonumber \\
&  \ \ \ \ \ \ \ \ \ \ \ \ - [(x_k- x'_{\pi(l)})^2]^{2- D/2}  -[(x'_{\pi (k)} - x_l)^2]^{2-D/2} \nonumber \\ 
& \ \ \ \ \ \ \ \ \ \ \ \ \left. + [(x'_{\pi (k)} - x'_{\pi (l)} )^2]^{2-D/2}\right\}\,.
\end{align}
We highlight that this does not depend on the spinor degrees of freedom, since their interaction is already gauge invariant, and expanding about $D = 4-2\epsilon$, at leading order it involves only conformal cross ratios of the $N$ worldline endpoints. Hence this is the same as in scalar QED \cite{LKFTWL1} and so the quenched approximation of the QED is sufficient to derive the LKFt, since $\Delta _\xi S_{i \pi}^{(k,l)}$ is zero for a virtual photon attached to a closed fermion loop ($x_l = x'_{\pi (l)}$). Then, the product of kernels transforms as
\begin{align}
&\braket{K^{x'_{\pi(1)}x_1} \cdots K^{x'_{\pi(n)}x_n} }_{\bar{A}, \xi+ \Delta \xi} \nonumber \\
& \ \ \ \ \ \ \ \  = \braket{K^{x'_{\pi(1)}x_1} \cdots K^{x'_{\pi(n)}x_n}}_{\bar{A}, \xi} \text{e}^{-\sum ^n_{k,l} \Delta _{\xi} S_{i \pi}^{(k,l)}},
\label{eqKTransform}
\end{align}
where the exponential can be factorised since it only depends on the endpoints of the lines. This transformation shares the multiplicative form of the original LFKt.

The next step is to include the pre-factors of ${[m + i \slashed{D}'_i -e \bar{A}_i]}$.  In \cite{PhysRevD.104.025014} it was found that the partial derivatives of the exponential in (\ref{eqKTransform}) from transforming the kernels cancel with other terms coming from the expected values involving insertions of $\bar{A}$.
So at the end, the transformation rule which arises from this analysis is
\begin{align}
&\mathcal{S}_\pi ^{\text{LKF}} (x_1 \dots x_n ; x' _{\pi (1)} \dots x' _{\pi (n)} | \xi + \Delta \xi) \nonumber \\
& \ \ = \text{e}^{- \sum ^n_{k,l} \Delta _{\xi} S_{i \pi} ^{(k,l)}} \mathcal{S}_\pi (x_1 \dots x_n ; x' _{\pi (1)} \dots x' _{\pi (n)} | \xi),
\end{align}
where the label ``LKF'' indicates that the LHS involves the $N$-point amplitude in the gauge $\xi + \Delta \xi$ plus extra gauge dependent parts of higher order diagrams generated by the exponent acting on the original amplitude.

\vspace{-0.5em}
\section{Conclusion}
Since $\sum ^n_{k,l} \Delta _{\xi} S_{i \pi} ^{(k,l)}$ is independent of the permutation, i.e., equal for each partial amplitude, the identical transformation rule is valid for the total amplitude, which defines the complete, generalized LKFt for spinor QED. It has turned out to be the same as in the scalar case (the original LKFt are recovered for $N=2$). In ongoing work these transformations are being analysed in the context of the Schwinger model, extended to propagation in external electromagnetic fields and applied to derive the analogous transformations of the interaction vertex.


\vspace{-1.5em}
\bibliography{mupbsamp}





\end{document}